\documentclass[twocolumn,prb,showpacs,preprintnumbers,amsmath,amssymb]{revtex4-1}

\usepackage{graphicx}
\usepackage{dcolumn}
\usepackage{bm}
\usepackage{verbatim}

\begin{document}


\title{Observation of anisotropic diamagnetism above the superconducting transition in iron-pnictide Ba$_{1-x}$K$_x$Fe$_2$As$_2$ single crystals due to thermodynamic fluctuations}

\author{J.~Mosqueira}
 \email{j.mosqueira@usc.es}
\author{J.D.~Dancausa}
\author{F.~Vidal}
\affiliation{LBTS, Facultade de F\'isica, Universidade de Santiago de Compostela, E-15782 Santiago de Compostela, Spain}
\author{S.~Salem-Sugui,~Jr.}
\affiliation{Instituto de Fisica, Universidade Federal do Rio de Janeiro, 21941-972 Rio de Janeiro, RJ, Brazil}
\author{A.D.~Alvarenga}
\affiliation{Instituto Nacional de Metrologia Normaliza\c c\~ao e Qualidade Industrial, 25250-020 Duque de Caxias, RJ, Brazil}
\author{H.-Q.~Luo}
\author{Z.-S.~Wang}
\author{H.-H.~Wen}
\affiliation{National Laboratory for Superconductivity, Institute of Physics and national Laboratory for Condensed Matter Physics, P.O. Box 603, Beijing 100190, People's Republic of China}

\date{\today}

\begin{abstract}
High resolution magnetization measurements performed in a high quality Ba$_{1-x}$K$_x$Fe$_2$As$_2$ single crystal allowed to determine the diamagnetism induced above the superconducting transition by thermally activated Cooper pairs. These data, obtained with magnetic fields applied along and transverse to the crystal $ab$ layers, demonstrate experimentally that the superconducting transition of iron pnictides may be explained at a phenomenological level in terms of the Gaussian Ginzburg-Landau approach for three-dimensional anisotropic superconductors.
\end{abstract}

\maketitle

\section{Introduction}

The recent discovery of superconductivity in iron-pnictides \cite{kamihara} has generated an intense research activity, in part because they provide an unexpected and very interesting scenario to study the interplay between magnetism and superconductivity.\cite{reviews} The interest for these compounds is enhanced by the fact that they share many properties with the high-$T_c$ cuprates (HTSC), as the layered structure, a similar evolution of the superconducting parameters with doping, and the proximity to a magnetic transition.\cite{reviews} 
As it is still the case of the HTSC, the pairing mechanism in superconducting iron pnictides is not yet known, making the phenomenological descriptions of their superconducting transition a central issue of the present researches on these compounds.
As first established in the pioneering experiments by M. Tinkham and coworkers in low-$T_c$ metallic superconductors,\cite{gollub} a powerful tool to probe at a phenomenological level the own nature of a superconducting transition is the diamagnetism induced above $T_c$ by thermally activated Cooper pairs.\cite{tinkham} The fluctuation diamagnetism above $T_c$ (sometimes called \textit{precursor diamagnetism}) was also early used to characterize at a phenomenological level the superconducting transition in both optimally-doped \cite{johnston,reviewHTSC} and underdoped \cite{carballeira} high-$T_c$ cuprates.  However, in spite of its interest the precursor diamagnetism received little attention due to both the relative smallness of the superconducting fluctuation effects above $T_c$ in these compounds and, mainly, to the need of high quality single crystals, with a relatively sharp superconducting transition.\cite{flucpnictides}

In this paper we report measurements in superconducting iron-pnictides of the fluctuation-induced diamagnetism above $T_c$, which were possible by applying high resolution magnetometry to a Ba$_{1-x}$K$_x$Fe$_2$As$_2$ single crystal with an excellent chemical and structural quality. These data allow an experimental demonstration that the superconducting transition of these materials may be explained at a phenomenological level in terms of the Gaussian Ginzburg-Landau approach for three-dimensional anisotropic superconductors (3D-AGL). This conclusion excludes phase incoherent superconductivity above $T_c$ in iron-pnictides, a long standing but still at present debated issue of the HTSC physics.\cite{citalarga,kondo}  Also, it provides a direct check of the fluctuations dimensionality: while bulk low-$T_c$ superconductors (LTSC) behave as three-dimensional (3D) and most HTSC as two-dimensional (2D), \cite{tinkham,reviewHTSC, carballeira} it has been recently proposed that iron-pnictides are in an intermediate regime as a consequence of their moderate anisotropy.\cite{tesanovic} 
In relation with this, another interesting aspect of our work is that the fluctuation diamagnetism is probed with magnetic fields applied along the two main crystallographic directions. This allows to accede directly to the superconducting anisotropy factor, but also to study an interesting issue recently observed in other moderately anisotropic superconductor (NbSe$_2$), as it is the anisotropy of non-local electrodynamic effects.\cite{NbSe2}

\section{Experimental details and results}

\subsection{Sample fabrication and characterization}

%
%
\begin{figure}[b]
\includegraphics[scale=.6]{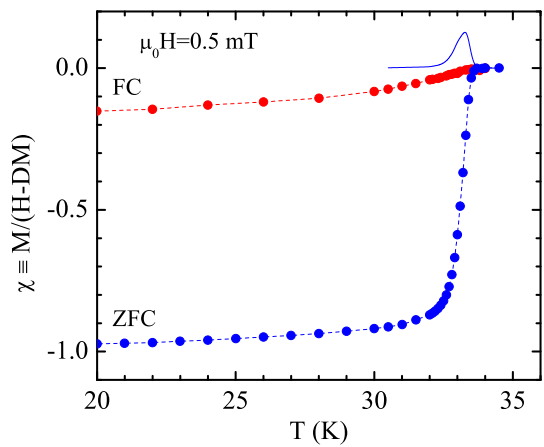}
\caption{(Color online) Temperature dependence of the FC and ZFC magnetic susceptibility (already corrected for demagnetizing effects) obtained with a 0.5 mT magnetic field perpendicular to the $ab$ layers. $T_c$ and its uncertainty were estimated from the maximum of $d\chi^{\rm ZFC}(T)/dT$ (solid line), and from the high-temperature half-width at half-maximum (this eludes the possible widening associated to the transit through the mixed state).}
\label{backs}
\end{figure}

The Ba$_{1-x}$K$_x$Fe$_2$As$_2$ sample (with $x=0.28$) used in this work is a $\sim1.1\times0.7\times 0.1$ mm$^3$ (430~$\mu$g) single crystal. Details of its growth procedure and characterization may be seen in Ref.~\onlinecite{growth}. We have chosen such a small crystal because it presents a well defined transition temperature which is essential to study critical phenomena. The measurements were performed with a magnetometer based in the superconducting quantum interference (Quantum Design, model MPMS-XL). To position the crystal with the magnetic field, $H$, applied parallel to the crystal $ab$ layers, we used a quartz sample holder (0.3 cm in diameter, 22 cm in length) to which the crystal was glued with GE varnish. Two plastic rods at the sample holder ends ($\sim0.3$ mm smaller than the sample space diameter) ensured that its alignment was better than $0.1^\circ$. For the measurements with $H\perp ab$ we made a groove ($\sim0.3$ mm wide) in the sample holder into which the crystal was glued also with GE varnish. 
The crystal alignment was checked by optical microscopy to be better than $5^\circ$. This allowed to determine the anisotropy factor from the anisotropy of the precursor diamagnetism with a $\sim0.5$\% uncertainty.\cite{errorgamma}

As a first magnetic characterization, in Fig.~1 it is presented the temperature dependence of the zero-field-cooled ZFC and field-cooled FC magnetic susceptibilities, measured with a magnetic field of 0.5 mT perpendicular to the $ab$ planes. These measurements were corrected for demagnetizing effects by using as demagnetizing factor $D=0.78$, which was obtained by approximating the crystal shape by the inscribed ellipsoid. A proof of the adequacy of this procedure is that the ZFC magnetic susceptibility in the Meissner region is close to the ideal value (-1) well below $T_c$. From these curves we estimated $T_c\approx33.2\pm0.2$~K, which attests the excellent quality of the sample, and allows to study the fluctuation-induced magnetization in almost the whole temperature range above $T_c$. Such a small transition width is in good agreement with the one determined from the resistive transition of crystals with the same composition and grown following the same procedure.\cite{growth}

\subsection{Measurements of the fluctuation diamagnetism around $T_c$ and background subtraction}

To measure the effect of superconducting fluctuations above $T_c$ (which according to the 3D-AGL approach, see below, is about $10^{-6}$-$10^{-7}$ emu), we used the \textit{reciprocating sample option}. It produces sinusoidal oscillations of the sample about the center of the detection system and improves the resolution by about two orders of magnitude with respect to the conventional DC option. Data were taken by waiting $\sim 3$ min for complete stabilization after the temperature was within 0.5\% the target temperature. For each temperature we averaged six measurements consisting of 15 cycles at 1 Hz frequency. The final resolution in magnetic moment, $m$, was in the $\sim10^{-8}$ emu range.

%
%
\begin{figure}[t]
\includegraphics[width=\columnwidth]{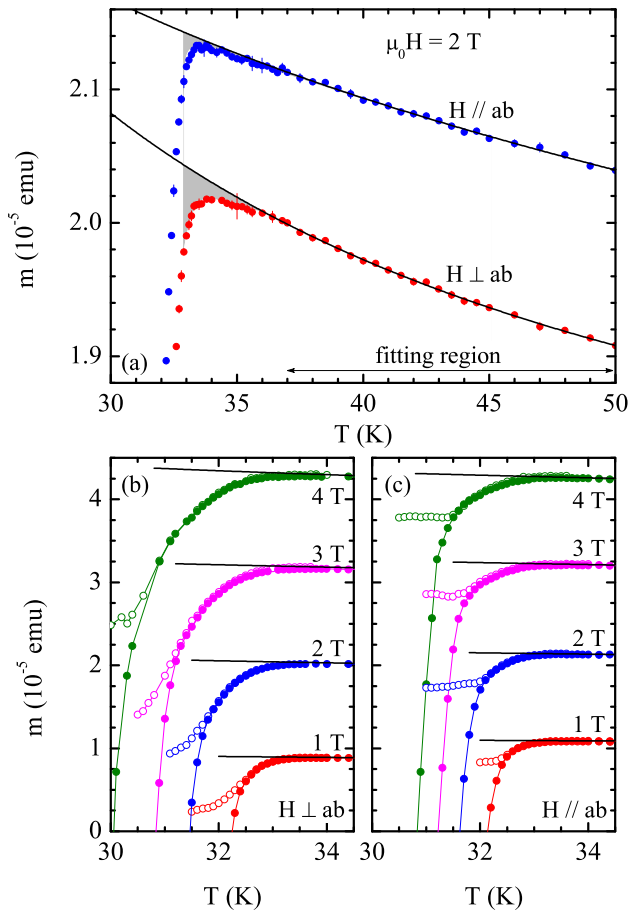}
\caption{(Color online) (a) Example of the temperature dependence up to 50 K of the as-measured magnetic moment for a magnetic field applied in the two main crystallographic directions. The normal-state backgrounds (lines) were determined by fitting Eq.~(\ref{back}) in the region indicated. The shaded areas represent the precursor diamagnetism. (b) and (c) Detail around the superconducting transition. Solid (open) symbols were obtained under ZFC (FC) conditions. The diamagnetism above $T_c$ is unobservable in this scale.}
\label{backs}
\end{figure}

An example of the as measured $m(T)$ data above $T_c$ is presented in Fig.~2(a). As may be clearly seen already in this figure, the $m(T)$ curves present a {\it rounding} just above $T_c$ (a factor $\sim 3$ larger in amplitude when $H \perp ab$), which extends few degrees above $T_c$. In view of the sharp low-field diamagnetic transition (Fig.~1) such a rounding cannot be attributed to $T_c$ inhomogeneities, and is a first evidence of the presence of an anisotropic precursor diamagnetism in these materials.  In the detail around $T_c$ presented in Figs.~2(b) and (c), it is clearly seen that the reversible region extends few degrees below $T_c$, allowing to study the \textit{critical} fluctuation regime, where $m(T)$ presents a similar anisotropy. 

The fluctuation-induced contribution was determined from the as measured $m(T)$ by subtracting the \textit{background} magnetic moment, $m_B(T)$, coming from the crystal normal-state and in a much lesser extent from the sample holder (this last contribution was estimated to be in the 10$^{-6}$ emu range). Such a background contribution was determined by fitting the function
\begin{equation}
m_B(T)=a+bT+\frac{c}{T}
\label{back}
\end{equation}
to the raw data in the temperature interval between 37 and 50 K ($a$, $b$ and $c$ are free parameters). The lower bound corresponds to a temperature above which the fluctuation-induced magnetic moment is expected to be smaller than the experimental uncertainty (see below). It is worth noting that the use of other plausible functions for $m_B(T)$, for instance $m_B(T)=a+b/(T-c)$, leads to imperceptible changes in the scale of Fig.~2(a).

\section{Data analysis}

\subsection{Theoretical background}

In view of the moderate anisotropy observed, the results of Fig.~\ref{backs} were analyzed in the framework of the 3D-AGL approach. This theory predicts that the fluctuation-induced magnetization for magnetic fields applied perpendicular or parallel to the crystal $ab$ layers ($M_\perp$ or $M_\parallel$, respectively) may be obtained from the result for 3D isotropic materials, $M_{\rm iso}$, through\cite{Klemm80,Hao92,Blatter92}
\begin{equation}
M_\perp(H)=\gamma M_{\rm iso}(H),
\label{transperp}
\end{equation}
and 
\begin{equation}
M_\parallel(H)=M_{\rm iso}\left(\frac{H}{\gamma}\right)=\frac{1}{\gamma}M_\perp\left(\frac{H}{\gamma}\right),
\label{transpara}
\end{equation}
where $\gamma$ is the superconducting anisotropy factor. 
This transformation was introduced by Klemm and Clem \cite{Klemm80} and generalized by Blatter \cite{Blatter92} and by Hao and Clem \cite{Hao92} to different observables and different regions in the $H-T$ phase diagram. In particular, it was shown to be also valid for the fluctuation region above $T_c$.\cite{Blatter92} 
For isotropic materials, the fluctuation magnetization above $T_c$ was calculated in Ref.~\onlinecite{PbIn} in the framework of a Gaussian GL approach including a \textit{total-energy} cutoff in the fluctuation spectrum.\cite{EPL_uncertainty} By combining that result with Eqs.~(\ref{transperp}) and (\ref{transpara}) it follows
\begin{eqnarray}
&&M_\perp=-\frac{k_BT\mu_0H\gamma\xi_{ab}(0)}{3 \phi_{0}^2}
\left[\frac{\arctan\sqrt{(c-\varepsilon)/\varepsilon}}{\sqrt{\varepsilon}}\right.\nonumber\\
&&\left.-\frac{\arctan\sqrt{(c-\varepsilon)/c}}{\sqrt{c}}
\right],
\label{Mperp}
\end{eqnarray}
and
\begin{equation}
M_\parallel=\frac{M_{\perp}}{\gamma^2},
\label{Mpara}
\end{equation}
where $\varepsilon=\ln(T/T_c)$ is the reduced temperature, $\xi_{ab}(0)$ the \textit{in-plane} coherence length amplitude, $c\approx0.55$ the cutoff constant,\cite{EPL_uncertainty} $k_B$ the Boltzmann constant, $\mu_0$ the vacuum magnetic permeability, and $\phi_0$ the magnetic flux quantum. These expressions are valid in the low magnetic field limit, i.e., for $H\ll\varepsilon H_{c2}^\perp(0)$ and, respectively, $H\ll\varepsilon H_{c2}^\parallel(0)=\varepsilon\gamma H_{c2}^\perp(0)$, where $H_{c2}^\perp(0)=\phi_0/2\pi\mu_0\xi_{ab}^2(0)$ is the upper critical field for $H\perp ab$ extrapolated linearly to $T=0$~K.
In absence of any cutoff (i.e., when $c\to\infty$) Eq.~(4) simplifies to 
\begin{equation}
M_\perp=-\frac{\pi k_BT\mu_0H\gamma\xi_{ab}(0)}{6\phi_{0}^2}\varepsilon^{-1/2},
\label{schmidt}
\end{equation}
which corresponds to the well known Schmidt result.\cite{Schmidts}

Equations (\ref{Mperp}) and (\ref{Mpara}) predict the vanishing of the precursor diamagnetism at $\varepsilon=c$ (i.e., $T\approx1.7T_c\approx56$~K in the present case). This has been experimentally confirmed in a number of HTSC and LTSC,\cite{EPL_uncertainty} but in the present case the small crystal size does not allow a quantitative check of this onset temperature.\cite{note}
The difference between Eq.~(\ref{Mperp}) and the conventional Schmidt approach  Eq.~(\ref{schmidt}) is dramatic at high reduced temperatures (typically above $\varepsilon=0.1$). However, it is still as high as 15\% for $\varepsilon$ as low as 0.01, which justifies the use of the total energy approach also at low reduced temperatures.

%
%
\begin{figure}[b]
\includegraphics[width=\columnwidth]{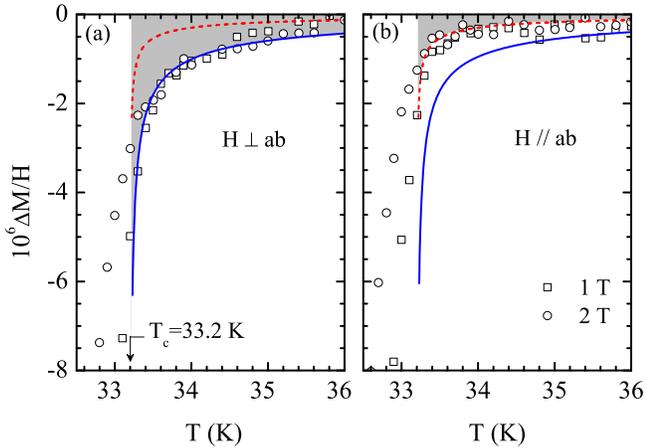}
\caption{(Color online) Detail around $T_c$ of the temperature dependence of the fluctuation magnetization (over $H$) for several magnetic fields applied in the two main crystallographic directions. The lines correspond to the Gaussian 3D-AGL approach in the low magnetic field limit (solid blue for $H\perp ab$ and dashed red for $H \parallel ab$). For a better comparison, both lines are presented in (a) and (b).}
\label{excess}
\end{figure}

\subsection{Fluctuation diamagnetism in the Gaussian region above $T_c$}

A detail of the fluctuation magnetization above $T_c$ (already corrected for the normal state contribution and normalized by the applied field) is presented in Fig.~3 for both $H\perp ab$ and $H\parallel ab$. In this representation we have only included magnetic fields up to 2 T, which are expected to be in the low magnetic field regime except in a narrow temperature interval close to $T_c$. As expected, both $M_\perp/H$ and $M_\parallel/H$ are independent of the applied magnetic field except very close to $T_c$, where finite-field effects are already observable.\cite{tinkham,PbIn,breakdown}  The best fit of Eqs.~(\ref{Mperp}) and (\ref{Mpara}) to the data in the low-field region is in good agreement with the temperature dependences and amplitudes observed. The values obtained for the two free parameters, $\xi_{ab}(0)=1.5$~nm (which leads to $\mu_0dH_{c2}^\perp/dT=-4.2$~T/K near $T_c$) and $\gamma=1.8$, are close to the ones in the literature.\cite{parametros} These values lead to a \textit{transverse} coherence length $\xi_c(0)=\xi_{ab}(0)/\gamma=0.83$~nm, larger than the Fe-layers periodicity length ($\sim$0.66~nm),\cite{growth} which justifies the applicability of the 3D-AGL approach. 
The $\xi_{ab}(0)$ value is well within the ones in HTSC, and $\gamma$ is just a few times smaller than in YBa$_2$Cu$_3$O$_{7-\delta}$. These similarities contrast with the differences between their precursor diamagnetism (allegedly unconventional in the HTSC).\cite{citalarga} Finally, $\xi_{ab}(0)$ is much shorter than in the moderately anisotropic 2H-NbSe$_2$, which could explain the non observation of non-local electrodynamic effects when $H\perp ab$ in the present case.\cite{NbSe2}

\subsection{Fluctuation diamagnetism in the critical region around $T_c$}

%
%
\begin{figure}[h]
\includegraphics[width=\columnwidth]{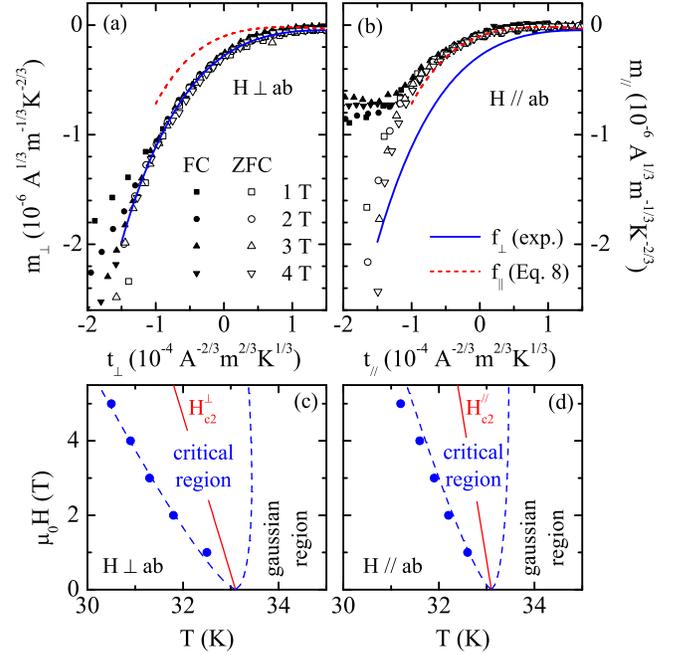}
\caption{(Color online) 3D-GL scaling of the magnetization in the critical region for $H\perp ab$ (a) and $H\parallel ab$ (b). The solid blue line is the \textit{experimental} scaling function for $H\perp ab$, while the dashed red one corresponds to $H\parallel ab$ and was calculated from $f_\perp$ through Eq.~(\ref{scalingfuncion}). (c) and (d) $H$-$T$ phase diagrams for $H\perp ab$ and, respectively, $H\parallel ab$. The corresponding $H_{c2}(T)$ lines were obtained from the superconducting parameters obtained in the analysis. The symbols are the lower bound of the reversible region, and the dashed lines the limits of the critical-region according to Eqs.~(\ref{criterioperp}) and (\ref{criteriopara}).}
\label{scaling}
\end{figure}

As a check of consistency of the above analysis we studied the data in the critical region, closer to the $H$-dependent critical temperature, where the fluctuations amplitude is so large that the Gaussian approximation breaks down.\cite{tinkham} This region is bounded by the so-called \textit{Ginzburg criterion},\cite{Ikeda} which according to the transformation to anisotropic materials may be written as
\begin{equation}
T\approx T_c^\perp(H)\pm T_c\left[\frac{4\pi k_B\mu_0H}{\Delta c\xi_c(0)\phi_0}\right]^{2/3}
\label{criterioperp}
\end{equation}
for $H\perp ab$, and  
\begin{equation}
T\approx T_c^\parallel(H)\pm T_c\left[\frac{4\pi k_B\mu_0H}{\Delta c\xi_c(0)\gamma\phi_0}\right]^{2/3}
\label{criteriopara}
\end{equation}
for $H\parallel ab$, where $\Delta c$ is the specific-heat jump at $T_c$. In this region, the 3D-GL approach in the lowest-Landau-level approximation predicts that $M_\perp(T,H)$ follows a scaling behavior, $m_\perp=f_\perp(t_\perp)$, the scaling variables being \cite{Ullah}
\begin{equation}
m_\perp\equiv \frac{M_\perp}{(HT)^{2/3}}
\label{perpvarm}
\end{equation}
and
\begin{equation}
t_\perp\equiv \frac{T-T_c^\perp(H)}{(HT)^{2/3}},
\label{perpvart}
\end{equation}
where $T_c^\perp(H)=T_c[1-H/H_{c2}^\perp(0)]$. According to Eq.~(\ref{transpara}), $M_\parallel(T,H)$ should scale as
\begin{equation}
m_\parallel=f_\parallel(t_\parallel)=\frac{f_\perp(t_\parallel\gamma^{2/3})}{\gamma^{5/3}}
\label{scalingfuncion}
\end{equation}
the scaling variables being now
\begin{equation}
m_\parallel\equiv \frac{M_\parallel}{(HT)^{2/3}}
\label{paravarm}
\end{equation}
and
\begin{equation}
t_\parallel\equiv\frac{T-T_c^\parallel(H)}{(HT)^{2/3}},
\label{paravart}
\end{equation}
where $T_c^\parallel(H)=T_c[1-H/H_{c2}^\parallel(0)]$, and $H_{c2}^\parallel(0)=\gamma H_{c2}^\perp(0)$.

In Figs.~\ref{scaling}(a) and (b) it is presented the scaling of the $M_\perp(T,H)$ and $M_\parallel(T,H)$ data according to Eqs.~(\ref{perpvarm}), (\ref{perpvart}), (\ref{paravarm}) and (\ref{paravart}). For that, $t_\perp$ and $t_\parallel$ were evaluated by using the $H_{c2}^\perp(0)$ and $\gamma$ values obtained in the analysis of the precursor diamagnetism in the Gaussian region, and $T_c=33.1$~K (in good agreement with the $T_c$ value resulting from Fig.~1). As may be clearly seen, both scalings are excellent and the relation between the observed scaling functions is in good agreement with Eq.~(\ref{scalingfuncion}). 
The applicability of the above scalings extends down to the lower bound of the reversible region. This is in agreement with Eqs.~(\ref{criterioperp}) and (\ref{criteriopara}), which when evaluated by using $\Delta c/T_c=0.1$ J/molK$^2$ (see Ref.~\onlinecite{zaanen}) and the superconducting parameters resulting from the previous analysis, are in an unexpected agreement with the observed lower bound of the reversible region for both $H$ directions [see Figs.~4(c) and (d)].
The 3D scaling was previously found to be adequate when $H\perp ab$, \cite{flucpnictides} but here its validity is extended to the case in which $H\parallel ab$.

\section{Conclusions}

The present experimental results and analysis represent a solid evidence that the precursor diamagnetism and the magnetization in the critical region around $T_c$ in Ba$_{1-x}$K$_x$Fe$_2$As$_2$ may be consistently explained at a phenomenological level in terms of a conventional GL approach for 3D anisotropic superconductors. This provides a strong constraint for any microscopic theory for the superconductivity in iron pnictides, and may have implications in systems like the HTSC, with similar superconducting parameters and possibly a similar mechanism for their superconductivity. It would be interesting to extend these measurements to other iron-pnictides with other type of substitution and doping levels, and also in larger samples to confirm whether the onset of the precursor diamagnetism is close to $1.7T_c$, as proposed by the total-energy cutoff scenario.

\section*{Acknowledgments}

This work was supported by the Spanish MICINN and ERDF \mbox{(grant no. FIS2010-19807)}, and by the Xunta de Galicia (grants no. 2010/XA043 and 10TMT206012PR). SSS and ADA acknowledge support from the CNPq and FAPERJ.



\begin{references} 

\bibitem{kamihara}Y. Kamihara , T. Watanabe, M. Hitano, and H. Hosono, J. Am. Chem. Soc. \textbf{130}, 3296 (2008).

\bibitem{reviews}For some reviews see, e.g., J. Paglione and R.L. Greene, Nature Phys. \textbf{6}, 645 (2010); K. Ishida, Y. Nakai, and H. Hosono, J. Phys. Soc. Jpn. \textbf{78}, 062001 (2009); D.C. Johnston, Advances in Physics \textbf{59}, 803, (2010).

\bibitem{gollub}J.P. Gollub, M.R. Beasley, R.S. Newbower, and M. Tinkham, Phys. Rev. Lett. \textbf{22}, 1288 (1969); J.P. Gollub, M.R. Beasley, R. Callarotti, and M. Tinkham, Phys. Rev. B \textbf{7}, 3039 (1973). 

\bibitem{tinkham}For an introductory review see, e.g., M. Tinkham, {\it Introduction to Superconductivity} (McGraw-Hill, New York, 1996), chap.~8. 

\bibitem{johnston}W.C. Lee, R.A. Klemm, and D.C. Johnston, Phys. Rev. Lett. \textbf{63}, 1012 (1989).

\bibitem{reviewHTSC}For a review see, e.g., F. Vidal, and M.V. Ramallo, \textit{The gap symmetry and fluctuations in high $T_c$ superconductors}, eds. J. Bok, G. Detscher, D. Pavuna, and S.A. Wolf (Plenum, London) 1998, p. 477.

\bibitem{carballeira}C. Carballeira, J. Mosqueira, A. Revcolevschi, and F. Vidal, Phys. Rev. Lett. \textbf{84}, 3157 (2000).

\bibitem{flucpnictides}The few works on the fluctuation diamagnetism in iron pnictides published until now focus only on the critical region below $T_c$. See, e.g., S. Salem-Sugui Jr., L. Ghivelder, A.D. Alvarenga, J.L. Pimentel Jr., H. Luo, Z. Wang, and H-H. Wen, Phys. Rev. B \textbf{80}, 014518 (2009); C. Choi, S. Hyun Kim, K-Y. Choi, M-H. Jung, S-I. Lee, X.F. Wang, X.H. Chen, and X.L. Wang, Supercond. Sci. Technol. \textbf{22}, 105016 (2009).

\bibitem{citalarga}See e.g., E. Bernardi, A. Lascialfari, A. Rigamonti, L. Roman\`o, M. Scavini, and C. Oliva, Phys. Rev. B \textbf{81}, 064502 (2010), and references therein. Note, however, that conventional GL approaches were shown to be in excellent quantitative agreement with the precursor diamagnetism of high quality samples of several HTSC families, see e.g., J. Mosqueira, C. Carballeira, M.V. Ramallo, C. Torr\'on, J.A. Veira, and F. Vidal, Europhys. Lett. \textbf{53}, 632 (2001); L. Cabo, J. Mosqueira, and F. Vidal, Phys. Rev. lett. {\bf 98}, 119701 (2007); J. Mosqueira, L. Cabo, and F. Vidal, Phys. Rev. B \textbf{76}, 064521 (2007); J. Mosqueira and F. Vidal, Phys. Rev. B \textbf{77}, 052507 (2008). Also, some of the anomalies observed in the precursor diamagnetism may be easily explained in terms of $T_c$ inhomogeneities, which in the case of doped HTSC could even have an \textit{intrinsic} origin, see e.g., L. Cabo, F. Soto, M. Ruibal, J. Mosqueira, and F. Vidal, Phys. Rev. B \textbf{73}, 184520 (2006); L. Cabo, J. Mosqueira, and F. Vidal, Phys. Rev. B \textbf{80}, 214527 (2009).

\bibitem{kondo}See, e.g., T. Kondo \textit{et al.}, Nature \textbf{457}, 297 (2009), and references therein.

\bibitem{tesanovic}J.M. Murray and Z. Te\u{s}anovi\'c, Phys. Rev. Lett. \textbf{105}, 037006 (2010).

\bibitem{NbSe2}F. Soto, H. Berger, L. Cabo, C. Carballeira, J. Mosqueira, D. Pavuna, and F. Vidal, Phys. Rev. B \textbf{75}, 094509 (2007).

\bibitem{growth}H. Luo, Z. Wang, H. Yang, P. Cheng, X. Zhu and H-H. Wen, Supercond. Sci. Technol. \textbf{21}, 125014 (2008).

\bibitem{errorgamma}In the presence of a crystal misalignment ($\theta_\perp$ or $\theta_\parallel$ when measuring with $H\perp ab$ or $H\parallel ab$, respectively) the ratio $M_\perp/M_\parallel$  is given by  $\gamma_{\rm eff}^2=(\gamma^2\cos^2\theta_\perp+\sin^2\theta_\perp)/(\gamma^2\sin^2\theta_\parallel+\cos^2\theta_\parallel)$ according to the 3D-AGL approach [see also J.~Mosqueira, M.V. Ramallo, A. Revcolevschi, C. Torr\'on, and F. Vidal, Phys. Rev. B \textbf{59}, 4394 (1999)]. By using $\theta_\perp\sim5^\circ$, $\theta_\parallel\sim0.1^\circ$, and $\gamma\sim 1.8$, the resulting $\gamma_{\rm eff}$ value is only $\sim 0.5$\% away from $\gamma$.

\bibitem{Klemm80}R.A. Klemm and J.R. Clem, Phys. Rev. B {\bf 21}, 1868 (1980).

\bibitem{Blatter92}G. Blatter, V.B. Geshkenbein, and A.I. Larkin, Phys. Rev. Lett. {\bf 68}, 875 (1992).

\bibitem{Hao92}Z. Hao and J.R. Clem, Phys. Rev. B {\bf 46}, 5853 (1992).

\bibitem{PbIn}J. Mosqueira, C. Carballeira, and F. Vidal, Phys. Rev. Lett. {\bf 87}, 167009 (2001); see also, J. Mosqueira, C. Carballeira, S.R. Curr\'as, M.T. Gonz\'alez, M.V. Ramallo, M. Ruibal, C. Torr\'on, and F. Vidal, J. Phys.: Condens. Matter \textbf{15}, 3283 (2003).

\bibitem{EPL_uncertainty}F. Vidal, C. Carballeira, S.R. Curr\'as, J. Mosqueira, M.V. Ramallo, J.A. Veira and J. Vina, Europhys. Lett. \textbf{59}, 754 (2002).

\bibitem{Schmidts}H. Schmidt, Z. Phys. {\bf 216}, 336 (1968); A. Schmid, Phys. Rev. {\bf 180}, 527 (1969).

\bibitem{note}According to Eq.~(\ref{Mperp}), the sample volume needed to determine the onset of the precursor diamagnetism with a 10\% accuracy in reduced temperature is 
$V\sim50\phi_0^2\delta m/k_BT_c\mu_0H\gamma\xi_{ab}(0)$, where $\delta m\sim10^{-8}$ emu is our experimental resolution in magnetic moment. In the present case this leads to $V\approx2$ mm$^3$ ($\sim 30$ times larger than in our crystal). In practice, the uncertainty associated to the background determination leads to an even more restrictive condition.

\bibitem{breakdown}F.~Soto, C. Carballeira, J. Mosqueira, M.V. Ramallo, M. Ruibal, J.A. Veira, and F. Vidal, Phys. Rev. B {\bf 70}, 060501(R) (2004).

\bibitem{parametros}U. Welp, R. Xie, A.E. Koshelev, W.K. Kwok, H.Q. Luo, Z.S. Wang, G. Mu, and H.H. Wen, Phys. Rev. B {\bf 79}, 094505 (2009). See also V.G. Kogan, Phys. Rev. B {\bf 80}, 214532 (2009), and references therein.

\bibitem{Ikeda}R. Ikeda, T. Ohmi, and T. Tsuneto, J. Phys. Soc. Jpn. {\bf 58}, 1377 (1989); \textit{ibid.} {\bf 59}, 1397 (1990); D.H. Kim, K.E. Gray, and M.D. Trochet, Phys. Rev. B {\bf 45}, 10801 (1992).

\bibitem{Ullah}S. Ullah and A.T. Dorsey, Phys. Rev. Lett. \textbf{65}, 2066 (1990); Phys. Rev. B \textbf{44}, 262 (1991).

\bibitem{zaanen}J. Zaanen, Phys. Rev. B \textbf{80}, 212502 (2009).

\end{references}
\end{document}